\documentclass[aps,amsfonts,amsmath,prd,preprint,nofootinbib]{revtex4-1}
\pdfoutput=1
\usepackage{graphicx}
\usepackage{latexsym}

\newcommand{\beq}{\begin{equation}}
\newcommand{\eeq}{\end{equation}}
\newcommand{\bea}{\begin{eqnarray}}
\newcommand{\eea}{\end{eqnarray}}

\begin{document}

\title{Inflation with negative potentials and the signature reversal symmetry}

\author{Audrey T. Mithani and Alex Vilenkin}
\address{
Institute of Cosmology, Department of Physics and Astronomy,\\ 
Tufts University, Medford, MA 02155, USA}

\begin{abstract}

We discuss recent papers by Hartle, Hawking and Hertog, which proposed that quantum cosmology could predict an inflating universe in models with negative scalar potentials.  Here we show that this is a result of an underlying symmetry which is manifest in both classical and quantum theory.  Moreover, we show that inflating solutions in such models generally develop tachyonic and ghost instabilities and are therefore unphysical in realistic theories.

\end{abstract}

\maketitle

\section{Introduction}

Hartle, Hawking and Hertog (HHH) have argued in \cite{H3,H3scalar} that, through quantum cosmology, a theory with a negative cosmological constant, $\Lambda < 0$, can describe a de Sitter universe with a positive effective cosmological constant.  They have reached this conclusion by studying the asymptotic behavior of the wave function of the universe $\Psi$ in the semiclassical regime for a de Sitter minisuperspace model.  An unusual feature of their solution for $\Psi$ is that it uses the signature of the spatial metric which is opposite to the standard choice.  (For a minisuperspace model this corresponds to using imaginary values of the scale factor.)  HHH argue, however, that this is not a problem, since the signature of the metric is not an observable quantity.  They have also extended the analysis to include a scalar field with a potential, as well as perturbations about the minisuperspace.  The main conclusion remained the same: models with slowly varying negative potentials have inflationary solutions that one expects to find in models with positive potentials.

In the present paper, we point out that the above conclusions follow from an underlying symmetry of the model, which we call the signature reversal symmetry.  This symmetry is present in both classical and quantum theory and is not restricted to minisuperspace.
Hence, one does not need to rely on quantum cosmology to obtain de Sitter solutions in $\Lambda < 0$ models: they are already present at the classical level.  We find, however, that the reversal of metric signature tends to make these solutions unstable.  In particular, it appears that instability cannot be avoided in models including vector fields.

We begin in the next section with a FRW minisuperspace model and outline how HHH obtained their de Sitter solutions for $\Lambda < 0$.  The classical and quantum versions of the signature reversal symmetry and their implications are discussed in Section 3. Finally, the instabilities of the signature-reversed solutions are analyzed in Section 4.

\section{Inflation with $\Lambda <0$}

The key HHH result can be illustrated in a simple minisuperspace model described by the action
\beq
{\cal I} = \int d^4 x \sqrt{-g} \left(\frac{R}{16\pi} - \rho_v\right),
\label{model1}
\eeq
where $R$ is the curvature scalar, $\rho_v$ is a constant vacuum energy density and the universe is assumed to be homogeneous, isotropic and closed:
\beq
ds^2 = \sigma^2 \left[ N^2(t)dt^2 - a^2(t)d\Omega_3^2\right].
\label{metric}
\eeq
Here, the factor $\sigma^2 = 2/ 3\pi$ is included to simplify later calculations,
$d\Omega_3^2$ is the metric on a unit 3-sphere, and we use Planck units where $G=c={\hbar}=1$.  

With the ansatz (\ref{metric}), the action takes the form  
\beq
{\cal{I}} = \int dt \frac{N}{2} \left[ a - \frac{a \dot{a}^2}{N^2} - \Lambda a^3 \right]
       \equiv \int {\cal L} dt,
\label{action1}
\eeq
where $\Lambda \equiv (4/3)^2 \rho_v$.  The canonical momentum conjugate to the scale factor $a$ is
\beq
p_{a} = \frac{d{\cal L}}{d{\dot a}} = -\frac{a\dot{a}}{N}.
\eeq
In the classical theory, variation of (\ref{action1}) with respect to the lapse function $N$ gives the Hamiltonian constraint ${\cal H}=0$, where
\beq
{\cal H}=-\frac{1}{2a}\left[p_a^2 +a^2(1-\Lambda a^2)\right].
\label{H}
\eeq
In quantum cosmology, the momentum $p_a$ is replaced by the differential operator $- i \partial / \partial a$, and the Hamiltonian constraint becomes the Wheeler-deWitt (WDW) equation
\beq
{\cal{H}}\Psi = 0.
\eeq
The ordering of operators $a$ and $p_a$ is unimportant in the semiclassical regime; hence we can write
\beq
\left[\frac{d^2}{d a^2} - U(a)\right] \Psi(a)=0,
\label{WDW}
\eeq
where
\beq
U(a)=a^2 - \Lambda a^4.
\label{U}
\eeq

In the semiclassical approximation,  solutions of the WDW equation (\ref{WDW}) can be obtained in the WKB form,
\beq
\Psi(a) = e^{i{\cal{S}}(a)},
\eeq
where $S(a)$ satisfies the classical Hamilton-Jacobi equation
\beq
\left(\frac{d{\cal S}}{da}\right)^2 + U(a)=0.
\label{HJ}
\eeq
Classical evolution, corresponding to the integral curves of the action ${\cal S}(a)$, can then be found from
\beq
\frac{da}{dt} = -\frac{N}{a} \frac{d{\cal{S}}}{da}.
\label{dadt}
\eeq
In order for $a(t)$ and $N(t)$ to be real, $d{\cal S}/da$ should also be real; then it follows from (\ref{HJ}) that the semiclassical regime is limited to the range where $U(a)\leq 0$.  For $\Lambda <0$, this range is empty, except for a single point $a=0$.  

Now, HHH suggest that a classical regime can be found even for $\Lambda < 0$, if the wave function $\Psi(a)$ is extended into the complex plane of $a$.  This new regime corresponds to purely imaginary values of $a=ib$ with $b$ real.  The 3-metric is then given by
\beq
ds_3^2 = +\sigma^2 b^2(t)d\Omega_3^2.
\label{metric1}
\eeq
Thus the signature of the metric has changed.  The wave function still has the WKB form,
\beq
\Psi(b) = e^{i{\tilde{\cal{S}}}(b)},
\label{Psib}
\eeq
with ${\tilde{\cal{S}}}$ satisfying the Hamilton-Jacobi equation
\beq
\left(\frac{d{\tilde{\cal S}}}{db}\right)^2 + U(b)=0,
\label{HJ1}
\eeq
where
\beq
U(b)=b^2 + \Lambda b^4.
\label{U1}
\eeq
The main difference from the previous case is that the classically allowed range $U(b)<0$ is now non-empty for $\Lambda <0$; it is $b>|\Lambda|^{-1/2}$.  The WKB solutions in this range are given by (\ref{Psib}) with
\beq
{\tilde{\cal S}}(b) = \pm \int^b db' \sqrt{-U(b')}. 
\eeq

With the substitution $a=ib$, the classical evolution equation (\ref{dadt}) becomes
\beq
i\frac{db}{dt} = \frac{N}{b} \frac{d{\tilde{\cal{S}}}}{db}.
\label{dbdt}
\eeq
Since $b$ and ${\tilde{\cal S}}$ are real, it is clear that $N$ should now be imaginary, so the time component of the metric also changes its sign to the opposite.  Hence, we define $N=i{\tilde N}$, where ${\tilde N}$ is real, and the 4-metric takes the form
\beq
ds^2 = -\sigma^2 \left[ {\tilde N}^2(t)dt^2 - b^2(t)d\Omega_3^2\right].
\label{4metric}
\eeq
The overall sign of the metric is opposite to the usual choice, but HHH point out that this sign is not an observable quantity, so the change of sign should not by itself be a problem.

The classical solution of Eq.~(\ref{dbdt}) for $b(t)$ depends on the choice of the lapse function ${\tilde N}(t)$.  For ${\tilde N}=1$, in which case the variable $t$ is the proper time along the comoving geodesics, we have
\beq
\frac{db}{dt}=\pm\sqrt{|\Lambda| b^2 - 1},  
\eeq
and the solution is 
\beq
b(t) =  |\Lambda|^{-1/2}\cosh\left[|\Lambda|^{1/2} (t-t_0)\right],
\label{dS}
\eeq
where $t_0$ is an arbitrary constant.
The metric (\ref{4metric}) with ${\tilde N}=1$ and scale factor (\ref{dS}) describes a Lorentzian de Sitter space of curvature radius $|\Lambda|^{-1/2}$, and HHH conclude that a universe with $\Lambda <0$ admits a semiclassical regime of de Sitter expansion.  This conclusion is entirely based on a semiclassical analysis, and one can expect that such de Sitter solutions should also exist in the classical theory.  We show in the following Section that this is indeed the case.

\section{The signature reversal symmetry}

Our key observation is that the classical equations of motion for the model (\ref{model1}) are invariant under the transformation
\beq
g_{\mu\nu}(x) \to -g_{\mu\nu}(x), ~~~~~~~~ \rho_v \to -\rho_v.
\label{symmetry}
\eeq
The scalar curvature changes sign under $g_{\mu\nu} \to -g_{\mu\nu}$, so the net result of the transformation is an overall change of sign of the action.  Hence, if $g_{\mu\nu}(x)$ is a solution extremizing the action with some value of vacuum energy density $\rho_v$, then $-g_{\mu\nu}(x)$ must be a solution for the value $-\rho_v$.  This is a general statement for the model (\ref{model1}); it does not assume any symmetry of the metric.

In the minisuperspace model (\ref{metric}), the symmetry transformation (\ref{symmetry}) reduces to 
\beq
a(t) \to ia(t), ~~~~~~ N(t) \to iN(t), ~~~~~~ \Lambda \to -\Lambda.
\label{symmetry1}
\eeq
The classical equation of motion (Hamiltonian constraint) for this model is
\beq
\left(\frac{\dot{a}}{N}\right)^2 + 1 - \frac{\Lambda}{3}a^2 = 0.
\label{eom}
\eeq
It is easily verified that this equation is invariant under (\ref{symmetry1}).

It follows from the symmetry (\ref{symmetry1}) that a reversed-signature FRW solution with $\Lambda < 0$ behaves like a regular solution with $\Lambda > 0$ that is, like de Sitter space.  This explains inflation with negative $\Lambda$ found by HHH.

This analysis can be extended to a more general class of models including scalar fields.  Consider, for example, 
\beq
{\cal I} = \int d^4 x \sqrt{-g} \left(\frac{R}{16\pi} + \frac{1}{2} g^{\mu\nu}\partial_\mu\phi \partial_\nu\phi -V(\phi)\right).
\label{model2}
\eeq
An appropriate symmetry transformation in this case is
\beq
g_{\mu\nu}(x) \to -g_{\mu\nu}(x), ~~~~~~~~ V(\phi) \to -V(\phi).
\label{symmetry2}
\eeq
Clearly, each term in the action (\ref{model2}) changes sign under this transformation. Hence, for any solution $\{g_{\mu\nu}(x),\phi(x)\}$ of the model with a potential $V(\phi)$, there is a solution $\{-g_{\mu\nu}(x),\phi(x)\}$ with a potential $-V(\phi)$.  In particular, slow-roll inflation can be obtained in models with slowly-varying negative potentials. We note that the symmetry (\ref{symmetry}) is a special case of (\ref{symmetry1}) for $V(\phi)=\rho_v = const$.  Note also that this symmetry can be trivially extended to models with several scalar fields.

An exact signature reversal symmetry is also present in quantum theory.  In quantum cosmology, the wave function of the universe is defined on the space of all 3-metrics $g_{ij}({\bf x})$ and 3-dimensional matter field configurations.  In the quantum version of the model (\ref{model2}), there is a single matter field $\phi({\bf x})$.  
The wave function $\Psi[g_{ij}({\bf x}),\phi({\bf x})]$ obeys the Wheeler-DeWitt equation \cite{DeWitt}
\beq
\left[ G_{ijkl}\frac{\delta}{\delta g_{ij}} \frac{\delta}{\delta g_{kl}} + \frac{\delta^2}{\delta\phi^2} + 2g\left(R^{(3)} - \frac{1}{2} g^{ij}\phi,_i \phi,_j - V(\phi)\right) \right] \Psi [g_{ij},\phi] = 0,
\label{WDWgeneral}
\eeq
where $g=\det(g_{ij})$ and
\beq
G_{ijkl}=\left( g_{ij}g_{kl}-g_{ik}g_{jl}-g_{il}g_{jk}\right).
\eeq
It can be easily verified that this equation is invariant under the transformation
\beq
g_{ij}({\bf x}) \to -g_{ij}({\bf x}), ~~~~~~~~ V(\phi) \to -V(\phi).
\label{symmetry3}
\eeq
This implies that if $\Psi[g_{ij},\phi]$ is a solution of Eq.~(\ref{WDWgeneral}) with some scalar potential $V(\phi)$, then $\Psi[-g_{ij},\phi]$ is a solution of (\ref{WDWgeneral}) with the potential $-V(\phi)$.

The symmetry (\ref{symmetry3}) of the Wheeler-DeWitt equation has been discussed earlier in Ref.~\cite{Vilenkin1988}, where it has been used to establish a relation between the Hartle-Hawking wave function of the universe $\Psi_{HH}$ \cite{HH} and the tunneling wave function $\Psi_T$ \cite{Vilenkin1986}.  The conjectured relation is
\beq
\Psi_{HH} = \Psi_T (g_{ij}\to -g_{ij}, V\to -V).
\eeq
It has been verified in \cite{Vilenkin1988} that this indeed holds in
the FRW minisuperspace version of the model
(\ref{model2}).\footnote{ As it stands, this relation is probably not
  applicable beyond minisuperspace, in view of the instabilities
  discussed in the next Section. Otherwise, one of the
  two wave functions might suffer from instabilities.  Note, however,
  that this issue is not very straightforward, because the definition
  of the Hartle-Hawking wave function involves some analytic continuations.}

The classical signature reversal symmetry was later discussed, in a different context, in \cite{Cvetic,Skenderis}. 

\section{Tachyons and ghosts}

In a scalar field model like (\ref{model2}) with a stable vacuum for the field $\phi$, the transformation $V(\phi)\to -V(\phi)$ will generally lead to an instability.  Consider for example a model with a negative vacuum energy density, $\rho_v <0$,  and a set of non-interacting scalar fields $\phi_j$ with masses $m_j$.  The corresponding scalar potential is
\beq
V(\phi) = \rho_v + \frac{1}{2} \sum_j m_j^2 \phi_j^2.
\label{Vphi}
\eeq
If the scalar fields are at the minima of their potentials, $\phi_j
=0$, then this is simply a model of pure gravity with a negative
$\rho_v$, and we know that it has a (signature-reversed) de Sitter solution that would
normally correspond to a positive vacuum energy density $|\rho_v|$.
However, if scalar fields are perturbed away from the vacuum, these
perturbations will now be described by the theory with potential
$-V(\phi)$.  So, if the potential (\ref{Vphi}) has stable minima with
$m_j^2 >0$, the perturbations about the de Sitter solution will have
tachyonic masses, $-m_j^2 < 0$, and will be unstable.  HHH have noted
this potential instability in their papers \cite{H3,H3scalar} and
suggested that if the present accelerated expansion is described by a
model with $\rho_v <0$, then the original model (\ref{Vphi}) must have
tachyonic scalar masses, so that the model with potential $-V(\phi)$
is stable.  We shall now point out another instability, which
arises in models including a vector field $A_{\mu}$.

The simplest model of this sort is
\beq
{\cal I} = \int d^4x \sqrt{-g} \left[ \frac{R}{16\pi} -\rho_v -\frac{1}{4}g^{\mu\sigma} g^{\nu\tau} F_{\mu\nu}F_{\sigma\tau} \right],
\eeq
where $F_{\mu\nu} = \partial_\mu A_{\nu} - \partial_\nu A_{\mu}$.  The first two terms in the square brackets change their sign under the transformation (\ref{symmetry}), while the last term does not.  The symmetry can be reinstated by adding a transformation for the vector field,\footnote{A similar symmetry transformation has been noted by D. Coule 
\cite{Coule} in the context of Schwarzschild-AdS solutions. We thank David Coule for bringing this paper to our attention.}$^,$\footnote{ If the vector field is coupled to other fields through the covariant derivative operator $D_\mu = \partial_\mu -ieA_\mu$, then (\ref{AiA}) should be accompanied by a transformation of the coupling, $e\to -ie$.}
\beq
A_\mu \to iA_\mu.
\label{AiA}
\eeq
But then the vector field action and its energy-momentum tensor change sign as a result of the transformation.  In other words, the vector field becomes a ghost, signaling an instability.

As one might expect, this problem persists in quantum cosmology.  The vector field may be treated in perturbative superspace, that is, as a perturbation about the de Sitter minisuperspace of Section 2.  
Following \cite{Louko}, the field may be expanded in spherical harmonics,
\begin{eqnarray}
A_0 & = & \sum_{nlm} r_{nlm} Q_{lm}^n \\
A_k & = & \sum_{nlmp} f_{nlmp} \left( S_k^{(p)} \right)^{n}_{lm} + \sum_{lm} g_{nlm} \left( P_k \right)^{n}_{lm} 
\end{eqnarray}
Here, $r_{nlm}$, $f_{nlmp}$ and $g_{nlm}$ are functions of $t$, $Q^n_{lm}$ are eigenfunctions of the scalar Laplacian on a unit 3-sphere, $\left(S_k^{(p)} \right)^{n}_{lm}$ and $\left(P_k \right)^{n}_{lm}$ are respectively the transverse and longitudinal eigenfunctions of the vector Laplacian.  The corresponding eigenvalues of the Laplacian depend only on the index $n$, which takes integer values, $n=1,2,3, ...$.  The indices $l$ and $m$ take values $l = 0,1,...,n-1$ and $m=-l, -l+1, ..., l$, and the parity index $p$ takes two possible values, $p=\pm 1$.  Hereafter, the labels $\{n,l,m,p\}$ will be denoted simply by $n$.

A convenient choice of gauge is $A_0 = 0, A_k^{|k} = 0$, where a vertical bar indicates a covariant derivative on a unit 3-sphere.  This corresponds to requiring $g_n=0, r_n=0$, and the action for the vector field becomes
\begin{eqnarray}
{\cal I}_{vec} & = & \sum_n {\cal I}_n \\
{\cal I}_n & = & \int dt \left[ \frac{a}{2N} \dot{f}^2_n - \frac{N}{2a}n^2 f_n^2 \right].
\end{eqnarray}
Now, the momentum conjugate to $f_n$ is
\beq
p_n = \frac{\partial{\cal{L}}}{\partial \dot{f_n}} = \frac{a\dot{f_n}}{N}
\eeq
and the Hamiltonian constraint is
\beq
2a{\cal{H}} = -p_a^2 - a^2 + \Lambda a^4 +\sum_n \left( p_n^2 + n^2f_n^2 \right) = 0.
\eeq
After quantization,  the Wheeler-DeWitt equation is
\beq
\left[ \frac{\partial^2}{\partial a^2} - a^2 + \Lambda a^4 - \sum_n \left( \frac{\partial^2}{\partial f_n^2} - n^2f_n^2 \right) \right] \Psi\left( a, f_n \right) =0.
\label{WDWvector}
\eeq

The sum in Eq.~(\ref{WDWvector}) is, up to an overall factor, the Hamiltonian of the vector field.  Different terms in this sum have the form of harmonic oscillator Hamiltonians and represent contributions of different vector field harmonics.  If we now substitute $a=ib$, as HHH did in Eq.~(\ref{WDW}), the resulting equation will have an opposite sign for $\Lambda$ {\it and} an opposite sign for the vector field Hamiltonian.  The latter change of sign indicates that the vector field has now become a ghost.

We thus conclude that signature-reversed solutions, such as the
inflationary solutions in models with $\Lambda
<0$ generally suffer from instabilities.  
If the original model with the standard metric signature is free of
tachyons and/or ghosts, then scalar fields become tachyons and vector
fields become ghosts upon signature reversal.  The
signature-reversed solutions are therefore unlikely to
describe realistic cosmologies.


Throughout this paper we focused on Lorentzian metrics (\ref{metric}), with $N$ and $a$ either both real or both imaginary.  Semiclassical solutions in the Euclidean regime can be obtained, e.g., for real $a$ and imaginary $N=i{\tilde N}$.  
In Refs.~\cite{H3,H3scalar}, HHH argue that, by a suitable choice of a complex lapse function $N(t)$, the wave function representing a signature-reversed inflating Lorentzian universe can be related to a wave function of a Euclidean, asymptotically Anti-de Sitter space.  The latter wave function can in turn be related to the partition function of a conformal boundary theory, using the AdS/CFT duality \cite{Maldacena}.  

This approach could provide a new method for calculating quantum probability distributions in the Lorentzian regime, assuming that the instabilities we discussed here can somehow be eliminated.  As we already mentioned, HHH suggest that scalar field instabilities can be removed by requiring that all scalar fields of the original theory are tachyons.  To deal with the vector field instability, one could similarly require that all vector fields of the original theory must be ghosts.  These fields would then be well behaved in the signature-reversed Lorentzian regime.  However, the action in the Euclidean regime would be unbounded from below, so the Euclidean theory, which is necessary to establish the AdS/CFT connection, would not be well defined.

\section*{Acknowledgements}

We are grateful to Jim Hartle, Thomas Hertog, Jaume Garriga and Ben Freivogel for useful discussions.
This work was supported in part by grants from the National Science Foundation and from the Templeton Foundation.

\end{document}